\documentclass[12pt]{article}

\usepackage{graphicx}
\usepackage[bookmarksnumbered,colorlinks,plainpages]{hyperref}

\begin{document}

\title{Power law distribution of R\'enyi entropy for equilibrium
systems having nonadditive energy}

\author{Qiuping A. Wang \\ Institut Sup\'erieur des Mat\'eriaux du Mans, \\
44, Avenue F.A. Bartholdi, 72000 Le Mans, France}

\date{}

\maketitle

\begin{abstract}
Using R\'enyi entropy, a possible thermostatistics for nonextensive systems is discussed. We
show that it is possible to have the $q$-exponential distribution function for nonextensive
systems having nonadditive energy but additive entropy. It is also shown that additive energy,
as a approximation within nonextensive statistics, is not suitable for discussing fundamental
problems for interacting systems.
\end{abstract}

{\small PACS : 02.50.-r, 05.20.-y, 05.70.-a}


\section{Introduction}

R\'enyi entropy\cite{Reny66}
\begin{equation}                                    \label{1}
S^R=\frac{\ln\sum_{i=1}^w p_i^q}{1-q},\;\;\;q\geq 0
\end{equation}
often applied in the study of multi-fractal and chaotic systems\cite{Beck1}, where $p_i$ is
the probability that the system is at the state labelled by $i$ (Boltzmann constant k=1) and
$w$ is the total number of the states. In these studies, R\'enyi entropy was associated with
the exponential probability distributions of BGS\cite{Beck1} which as a matter of fact are not
the distributions derived from this entropy. Since the proposal\cite{Tsal88} of the
nonextensive statistical mechanics (NSM), there is a growing interest in R\'enyi entropy which
has been compared to Tsallis entropy\cite{Tsal88} $S^T=\frac{\sum_{i=1}^w p_i^q-1}{1-q}$,
associated with a $q$-exponential distribution $exp_q(x)=[1+(1-q)x]^{1/(1-q)}$, in the
discussions of possible nonextensive statistics and the relative fundamental problems such as
thermodynamic equilibrium and
stability\cite{Curado,Ramshaw,Lesche,Abe1,Raggio,Wada,Abe3,Tsal02,Wang03} for systems having
additive energy or infinite number of states\cite{Lesche}.

Regarding the possible statistics of R\'enyi entropy, some questions should be asked : $S^R$
is additive just as, e.g., the entropies of Boltzmann-Gibbs-Shannon statistics (BGS) for
systems having product joint probability\cite{Wang03b}, but should it be associated to
independent systems having additive energy just as in BGS? Should it be associated to
exponential probability distributions as has been done in many works? Are there alternative
distributions intrinsic to R\'enyi entropy if it can be maximized to get thermodynamic
equilibrium? If the R\'enyi statistics is not extensive, what are its possible nonextensive
properties? Due to the importance of $S^R$ in the study of chaos and fractals and since $S^R$
is identical to Boltzmann entropy $S=\ln W$ for the fundamental microcanonical
ensemble\cite{Beck1}, the possible answers to the above questions would be interesting for
both BGS and its possible extensions based on $S^R$.

In this paper, I will present a thermo-statistics derived from R\'enyi entropy for equilibrium
systems with {\it nonadditive energy required by the existence of thermodynamic
equilibrium}\cite{Wang02a}. This formalism may be considered as an alternative to BGS and to
NSM for {\it canonical systems} with additive entropy but nonadditive energy\cite{Gross}.

\section{R\'enyi and Tsallis entropies}
There is a monotonic relationship between these two entropies :
\begin{equation}                                    \label{2}
S^R=\frac{\ln[1+(1-q)S^T]}{1-q} \;\;or\;\; S^T=\frac{e^{(1-q)S^R}-1}{1-q},
\end{equation}
and, for complete probability distribution ($\sum_{i=1}^wp_i=1$) in microcanonical ensemble,
$S^R$ is identical to the Boltzmann entropy $S$ :
\begin{equation}                                    \label{3}
S^R=S=\ln w
\end{equation}
since $\sum_{i=1}^wp_i^q=w^{1-q}$. Other properties of $S^R$ can be found in
\cite{Reny66,Beck1,Curado}.

The concavity of $S^R$ and $S^T$ for $q>1$ is shown in Figure 1 and 2, respectively (they are
convex and have minimums for $q<0$, so we do not consider this case in this paper). It is
worth noticing that the two entropies get their maximum at the same time for any $q$. But the
maximum of $S^R$ is $\ln W$ and independent of $q$, while the maximum of $S^T$ is
$\frac{W^{1-q}-1}{1-q}$ and decreases down to zero when $q\rightarrow \infty$ and increases up
to $W-1$ when $q\rightarrow 0$. It should be mentioned that $S^R$ is not always concave for
$q>1$, as shown in Figure 1.

\begin{figure}[ht] \label{f1}
\begin{center}
\includegraphics[width=4in,height=3in]{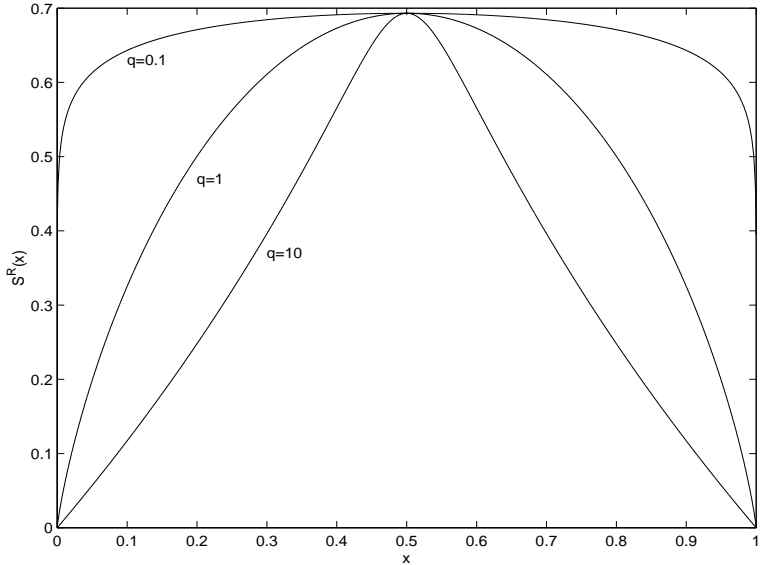}
\end{center}
\caption{The concavity of R\'enyi entropy $S^R$ for $q>0$ with a two probability
distribution $p_1=x$ and $p_2=1-x$. Note that the maximal value does not change with $q$.
The maximum becomes more sharp for larger $q$. The curve for $q>1$ shows that $S^R$ is
not always concave, but the maximum remains the unique extremum.}
\end{figure}

\begin{figure}[ht] \label{f2}
\begin{center}
\includegraphics[width=4in,height=3in]{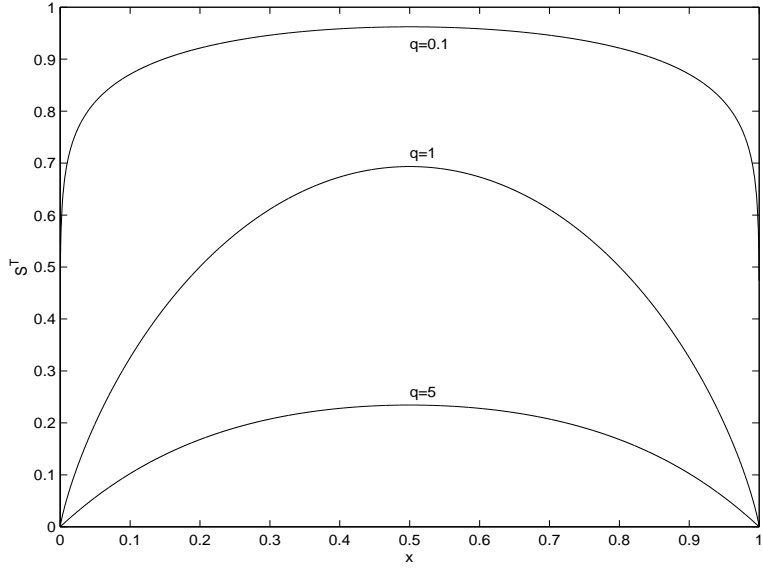}
\end{center}
\caption{The concavity of Tsallis entropy $S^T$ for $q>0$ with a two probability distribution
$p_1=x$ and $p_2=1-x$. Note that the maximal value increases from zero to unity with $q$
decreasing from infinity to zero.}
\end{figure}

Due to the fact that $S^R$ is a monotonically increasing function of $S^T$, they reach the
extremum together (see Figures 1 and 2). One can hope that the maximum entropy (for
$q>0$\cite{Curado}) will give same results with same constraints. Indeed, R\'enyi entropy has
been used to derive, by maximum entropy method, the Tsallis $q$-exponential distribution
within an additive energy formalism\cite{Lenzi,Bashkirov}. In what follows, we present a
thermostatistics based on $S^R$ for nonextensive {\it systems having additive entropy and
nonadditive energy.}.

\section{Canonical distribution of R\'enyi}
We suppose complete distribution $\sum_{i=1}^wp_i=1$ and $U=\sum_{i=1}^wp_iE_i$ where $U$
is the internal energy and $E_i$ the energy of the system at the state $i$. We will
maximize as usual the following functional :
\begin{equation}                                    \label{4}
F=\frac{\ln\sum_ip_i^q}{1-q}+\alpha\sum_{i=1}^wp_i-\gamma\sum_{i=1}^wp_iE_i.
\end{equation}
We get
\begin{equation}                                    \label{5}
p_i \propto [\alpha-\gamma E_i]^{1/(q-1)}.
\end{equation}
Since $S^R$ recovers Boltzmann-Gibbs entropy $S=-\sum_{i=1}^wp_i\ln p_i$ when $q=1$, it
is logical for us to require that Eq.(\ref{5}) recovers the conventional exponential
distribution for $q=1$. This leads to
\begin{equation}                                    \label{6}
p_i=\frac{1}{Z}[1-(q-1)\beta E_i]^{1/(q-1)}.
\end{equation}
where $(q-1)\beta=\gamma/\alpha$ and $Z=\sum_{i=1}^w[1-(q-1)\beta E_i]^{1/(q-1)}$. We
will show the physical meaning of $\beta$ later.

Note that the second derivative $\frac{d^2F}{dp_i^2}=
-\frac{qp_i^{q-2}}{\sum_ip_i^q}[1+\frac{q}{1-q}\frac{p_i^q}{\sum_ip_i^q}]$ is negative for any
distribution only when $q<1$. In general $\frac{d^2F}{dp_i^2}$ may be positive for $q>1$ so
that the above distribution Eq.(\ref{6}) is not stable.

This formalism can be applied to an important family of chaotic behaviors described by L\'evy
flight\cite{Alemany} if we assume $0\leq q\leq 2/3$. In this case, the first moment $\bar{x}$
calculated with the usual expectation is convergent for $1/2\leq q\leq 2/3$ and can be used as
a constraint of maximum entropy to obtain $p(x)\sim [1+(1-q)\beta x]^{-1/(1-q)}$. For L\'evy
flight distribution with large $x$, $p\sim x^{-1-\gamma}$, one obtain : $q=\gamma/(1+\gamma)$
($0<\gamma<2$).

\section{Mixte character : nonadditive energy and additive entropy}
It has been shown\cite{Wang02a} that, for thermal equilibrium to take place in nonextensive
systems, the internal energy of the composite system $A+B$ containing two subsystems $A$ and
$B$ must satisfy
\begin{equation}                                    \label{7}
U(A+B)=U(A)+U(B)+\lambda U(A)U(B)
\end{equation}
which means
\begin{equation}                                    \label{8}
E_{ij}(A+B)=E_i(A)+E_j(B)+\lambda E_i(A)E_j(B)
\end{equation}
where $\lambda$ is a constant. Applying Eq.(\ref{8}) to Eq.(\ref{6}), we straightforwardly get
the product joint probability :
\begin{equation}                                    \label{9}
p_{ij}(A+B)=p_i(A)p_j(B)
\end{equation}
and the additivity of $S^R$ :
\begin{equation}                                    \label{10}
S^R(A+B)=S^R(A)+S^R(B)
\end{equation}
if $\lambda=(1-q)\beta$. So $S^R$ is essentially different from $S^T$, because in this
case $S^T$ is nonadditive with $S^T(A+B)=S^T(A)+S^T(B)+(1-q)S^T(A)S^T(B)$. Note that we
do not need independent or noninteracting or weakly interacting subsystems for
establishing the additivity of $S^R$ or the nonadditivity of $S^T$, as discussed in
\cite{Wang02a,Wang02b,Wang02c}). So in this formalism, we can deal with interacting
systems with nonadditive energy but additive entropy.

\section{Zeroth law and temperature}
It is easy to show that, from Eq.(\ref{6}),
\begin{equation}                                    \label{11}
S^R=\ln Z+\ln[1+(1-q)\beta U]/(1-q).
\end{equation}
So we have
\begin{equation}                                    \label{12}
\beta= [1+(1-q)\beta U]\frac{\partial S^R}{\partial U}
\end{equation}
or
\begin{equation}                                    \label{13}
\frac{1}{\beta}=\frac{\partial U}{\partial S^R}-(1-q)U.
\end{equation}
Since $[1+(1-q)\beta U]$ is always positive ($q$-exponential probability cutoff), $\beta$
has always the same sign as $\frac{\partial S^R}{\partial U}$. $\beta$ can be proved to
be the effective inverse temperature if we consider the zeroth law of thermodynamics. Let
$\delta S^R$ be a small change of $S^R$ of the isolated composite system $A+B$.
Equilibrium means $\delta S^R=0$. From Eq.(\ref{10}), we have $\delta S^R(A)=-\delta
S^R(B)$. However, from Eq.(\ref{7}), the energy conservation of $A+B$ gives
$\frac{1}{[1+(1-q)\beta U(A)]}\delta U(A)=-\frac{1}{[1+(1-q)\beta U(B)]}\delta U(B)$.
That leads to
\begin{equation}                                    \label{14}
[1+(1-q)\beta U(A)]\frac{\partial S^R(A)}{\partial U(A)}=[1+(1-q)\beta
U(B)]\frac{\partial S^R(B)}{\partial U(B)}
\end{equation}
or $\beta(A)=\beta(B)$ which characterizes the thermal equilibrium.

\section{Some ``additive'' thermodynamic relations}
Due to the mixte character of this formalism with additive entropy and nonadditive
energy, all the thermodynamic relations become nonlinear. In what follows, we will try to
simplify this formal system and to give a linear form to this nonlinearity.

Using the same machinery as in \cite{Wang02c} which gives an extensive form to the
nonextensive Tsallis statistics, we define an additive deformed energy $E$ as follows :
\begin{equation}                                    \label{15}
E=\ln[1+(1-q)\beta U]/(1-q)\beta
\end{equation}
which is identical to $U$ whenever $q=1$. Note that $E(A+B)=E(A)+E(B)$. In this way,
Eq.(\ref{11}) can be recast into
\begin{equation}                                    \label{11a}
S^R=\ln Z+\beta E.
\end{equation}
So that $\beta=\frac{\partial S^R}{\partial E}$. The first law can be written as
\begin{equation}                                    \label{16}
\delta E=T\delta S^R+Y\delta X
\end{equation}
where $Y$ is the deformed pressure and $X$ the coordinates (volume, surface ...) and
$T=1/\beta$. The deformed free energy can be defined by

\begin{equation}                                    \label{17}
F=E-TS^R=-T\ln Z,
\end{equation}
so that $Y=\frac{\partial F}{\partial X}$. The real pressure is $Y^R=[1+(1-q)\beta U]Y$
and the work is $\delta W=Y^R\delta X$. The deformed heat is $\delta Q=T\delta S^R$ and
the real heat is $\delta Q^R=[1+(1-q)\beta U]\delta Q$

\section{Grand-canonical distributions}
It is easy to get the grand-canonical ensemble distribution given by

\begin{equation}                                        \label{18}
p_i=\frac{1}{Z}[1-(q-1)\beta(E_i-\mu N_i)]^{1/(q-1)},
\end{equation}
which gives
\begin{equation}                                    \label{19}
S^R=\ln Z+\ln[1+(1-q)\beta U]/(1-q)+\ln[1-(1-q)\beta\omega N]/(1-q).
\end{equation}
Let $M$ be the deformed particle number : $M=\ln[1-(1-q)\beta\omega N]/(1-q)\beta\omega$,
Eq.(\ref{19}) becomes
\begin{equation}                                    \label{11b}
S^R=\ln Z+\beta E+\beta\omega M.
\end{equation}
$M$ must be additive, so that $N$ is nonadditive satisfying
\begin{equation}                                    \label{20}
N(A+B)=N(A)+N(B)+(1-q)\beta\omega N(A)N(B)
\end{equation}
Due to the distribution function of Eq.(\ref{18}), the quantum distributions will be identical
to those in NSM\cite{Wang02d}.

\section{R\'enyi statistics with additive energy?}
Now we know that $S^R$ should be intrinsically associated with the $q$-exponential
distributions. Only when $q=1$ this statistics recovers BGS and the $q$-exponential becomes
the usual exponential function. Then an interesting question is whether or not this statistics
may be associated with additive energy when $q\neq 1$.

Supposing $A$ and $B$ are independent, i.e., $E_{ij}(A+B)=E_i(A)+E_j(B)$, let us see the
probability of the system $A+B$ for a joint state $ij$ :
\begin{eqnarray}                                  \label{21}
p_{ij}(A+B) &=& \frac{1}{Z(A+B)}[1-(q-1)\beta(E_i(A)+E_j(B))]^{1/(q-1)} \\ \nonumber &=&
p_i(A)p_{j\mid i}(B\mid A)
\end{eqnarray}
where
\begin{eqnarray}                                  \label{22}
p_i(A)=\frac{1}{Z(A)}[1-(q-1)\beta E_i(A)]^{1/(q-1)}
\end{eqnarray}
is the probability for $A$ to be at the state $i$ and
\begin{eqnarray}                                  \label{23}
p_{j\mid i}(B\mid A)=\frac{1}{Z_i(B\mid A)}[1-(q-1)\beta e_{j\mid i}(A\mid B)]^{1/(q-1)}
\end{eqnarray}
is a conditional probability for $B$ to be at a state $j$ with energy $e_{j\mid i}(A\mid
B)=E_j(B)/[1-(q-1)\beta E_i(A)]$ if $A$ is at $i$ with energy $E_i(A)$. In this case, the
total entropy is given by
\begin{eqnarray}                                  \label{24}
S^R(A+B) &=& \frac{\ln[\sum_ip_i(A)^q\sum_jp_{j\mid i}(B\mid A)^q]}{1-q} \\ \nonumber &\neq&
S^R(A)+S^R(B).
\end{eqnarray}
This is in contradiction with Eq.(\ref{10}). So {\it $S^R$ is no more additive with additive
energy}. As a matter of fact, Eq.(\ref{24}) is wrong because Eq.(\ref{21}) does not hold if we
consider the product probability Eq.(\ref{9}). We would get
\begin{eqnarray}                                  \label{25}
p_{j}(B) &=& \frac{1}{Z(B)}[1-(q-1)\beta E_j(B)]^{1/(q-1)} \\ &=& \frac{1}{Z_i(B\mid
A)}[1-(q-1)\beta e_{j\mid i}(A\mid B)]^{1/(q-1)}
\end{eqnarray}
which implies $E_j(B)=e_{j\mid i}(A\mid B)=E_j(B)/[1-(q-1)\beta E_i(A)]$ and is valid only
when $q=1$. This means that additive energy may force back the nonextensive statistics to BGS
if the product joint probability applies.

We would like to discuss in passing the problem of thermodynamic instability of R\'enyi
entropy which has been shown\cite{Lesche} to be non-observable and instable because an
arbitrarily small variation $\delta$ in probability distribution may lead to an important
variation in $S^R$ for a system having infinite number of states $w$. According to this
analysis, R\'enyi entropy can be physically useful for finite systems having finite number of
states.

However, it should be noted that the Lesche's conclusion for infinite $w$ is reached with the
asymptotic behavior of $S^R$ for the case of $w\rightarrow\infty$ and of finite variation
$\delta$ of probability distributions, i.e., $1/w$ is negligible compared to $\delta$. This is
a very harsh condition if we consider that $\delta$ must be arbitrarily small for
observability condition. It should be noted that the asymptotic behavior of $S^R$ for finite
$\delta$ and $w\rightarrow \infty$ is different from the asymptote for arbitrarily small
$\delta$ and arbitrary $w$ which can be very large, e.g., $\delta$ is smaller than or of same
order of magnitude as $1/w$. This second asymptotic behavior should be more general to our
opinion because it applies for any system. Taking the probability distributions proposed by
Lesche and making the same calculations without neglecting anything, one gets, for both $q>1$
and $q<1$, $\Delta S^R(\delta,w)/S_{max}\propto (\delta/2)^q$ for arbitrarily small
$\delta\rightarrow 0$. The observability condition\cite{Lesche} is satisfied. This result is
in addition consistent with the fact that $S^R$ is a monotonic function of $S^T$ which is
observable according to the same analysis\cite{Abe1}. We indeed have
$dS^R=\frac{dS^T}{1+(1-q)S^T} =\frac{dS^T}{\sum_ip_i^q}$. So if $dS^T/S^T\rightarrow 0$, we
also have $dS^R/S^R$ for finite $S^R$ and $S^T$.

\section{Conclusion}
In summary, the additive R\'enyi entropy is associated with nonadditive energy to give an
nonextensive thermostatistics characterized by $q$-exponential distributions which have been
proved to be useful for many systems showing non Gaussian and power law
distributions\cite{Tsalsite}. This formalism would be useful for interacting nonextensive
systems whose information and entropy may be additive or approximately additive. The problem
of the instability of Renyi entropy is reviewed. We think that this entropy may be physically
useful for any number of states.

A important point should be underlined following the result of the present work. R\'enyi
entropy is identical to Boltzmann one for microcanonical ensemble. So, theoretically, its
applicability to systems with nonadditive energy means that Boltzmann entropy may also be
applied to nonextensive microcanonical systems as indicated by Gross\cite{Gross}.

\section*{Acknowledgement}
The author thanks with great pleasure Professors S. Abe, D.H.E. Gross for valuable discussions
on some points of this work and for bringing my attention to important references.

\end{document}